\documentclass[sigconf]{acmart}
 \usepackage{amsmath,amssymb,amsfonts,bm}
\usepackage{graphicx}
\usepackage[most]{tcolorbox}
\usepackage{multicol} 
\usepackage{enumitem}
\usepackage{textcomp}
\usepackage{xcolor}
\usepackage{float}
\usepackage{tabularx}
\usepackage{balance}
\usepackage{marvosym}
\usepackage{multirow}
\usepackage{array}
\usepackage{float}  
\usepackage{enumitem}
\usepackage{booktabs}
\usepackage[ruled]{algorithm2e} 
\usepackage{pifont}
\usepackage{url}
\usepackage{algpseudocode}
\usepackage{subfig}
\usepackage[ruled]{algorithm2e}
\begin{document}
\title{Experience-Driven Online Learning for Agentic\\ Recommender Systems}

\author{Xin Xia}
\affiliation{%
	\institution{Griffith University}
	\city{Brisbane}
	\country{Australia}}
\email{xin.xia@griffith.edu.au}

\author{Junliang Yu}
\affiliation{%
	\institution{Griffith University}	
	\city{Brisbane}
	\country{Australia}}
\email{junliang.yu@griffith.edu.au}
\author{Min Gao}
\affiliation{%
	\institution{Chongqing University}	
	\city{Chongqing}
	\country{China}}
\email{gaomin@cqu.edu.au}


\author{Hongzhi Yin}
\affiliation{%
	\institution{The University of Queensland}
	\city{Brisbane}
	\country{Australia}}
\email{h.yin1@uq.edu.au}
\author{Shane Culpepper}
\affiliation{%
	\institution{The University of Queensland}
	\city{Brisbane}
	\country{Australia}}
\email{s.culpepper@uq.edu.au}

\begin{abstract}
Large language models have enabled a new generation of agentic recommender systems, where recommendation can be formulated as an interactive decision-making process. 
However, most existing approaches learn either from offline training on historical interaction data or through reward-guided optimization, making it difficult to learn continually from post-deployment interactions. 
This limitation is amplified in dynamic recommendation environments, where evolving user preferences and interaction patterns generate rich and structured feedback that is difficult to faithfully capture with scalar reward proxies alone. This raises a fundamental question: how can rich interaction trajectories be transformed into reusable knowledge for continual improvement? 
To address this issue, we propose  \textbf{O}nline \textbf{E}xperience-Driven \textbf{A}gentic \textbf{R}ecommendation (\textbf{OEAR}), which introduces experience as an intermediate form of reusable knowledge that bridges interaction and model improvement.
OEAR extracts transferable experiences from runtime agent interactions and internalizes them into agent backbones through on-policy experience distillation. Across successive learning rounds, updated agents generate new interaction trajectories that in turn provide supervision for subsequent improvement, forming a closed interaction–learning loop. Extensive experiments on three benchmarks show that OEAR consistently outperforms existing baselines in both recommendation and user simulation, with performance improving progressively over successive online learning rounds.
	\end{abstract}



	


\maketitle
\section{Introduction}
\begin{figure}[t]
    \centering
    \includegraphics[width=\columnwidth]{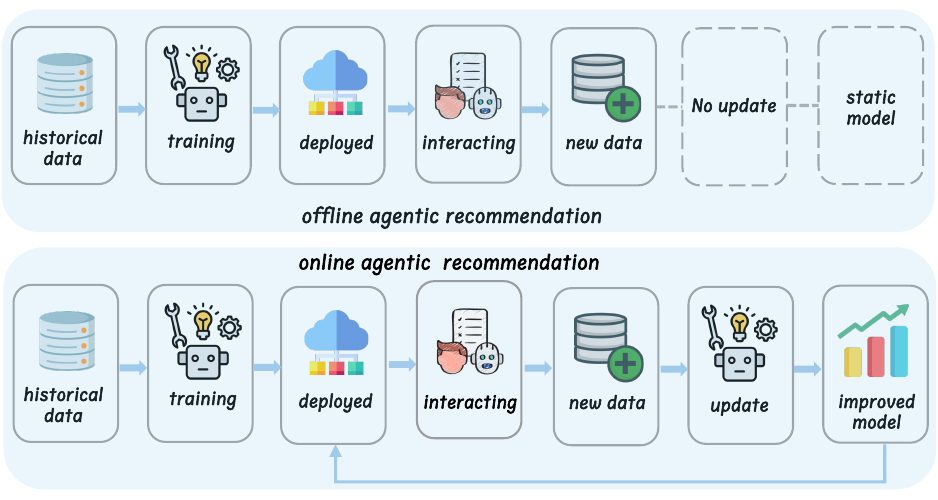}
    \vspace{-15pt}
    \caption{Online learning vs. Offline learning for agentic recommendation.}
    \label{figure1}
    \vspace{-15pt}
\end{figure}
Large language models \cite{wu2024survey, harte2023leveraging, zhao2024recommender} are giving rise to a new paradigm of agentic recommendation, in which recommendation is formulated not merely as a one-shot prediction problem, but as an interaction-driven decision process realized through multi-turn interactions and multi-agent collaborations \cite{zhang2024agentcf, cai2025agentic, wang2024macrec,maragheh2025future}.
In this paradigm, recommendation quality is determined not only by the ability to estimate user-item relevance, but also by how effectively the recommender system interprets user feedback, revises its decisions over time, and sustains coherent behavior across repeated interactions. 
The resulting shift is substantial: recommendation is no longer viewed solely as a static preference matching problem using historical data, but rather as a dynamic interaction process in which the recommender system must continually respond to user signals evolving over time \cite{huang2025recommender,huang2025towards}.

Despite reformulating the solution, the dominant learning paradigm remains largely unchanged. 
Existing recommender systems, including recent agentic variants, are still initially improved offline, either from fixed historical interaction data or through optimization using pre-collected or simulated trajectories \cite{wang2023recagent, cai2025agentic}.
This introduces a fundamental mismatch between how recommendation systems are deployed and how they are improved. 
At deployment time, the recommender system operates in a non-stationary environment, where user preferences, candidate spaces, and interaction patterns evolve over time. 
However, during training, model improvement remains largely aligned to the supervision and optimization signals specified before deployment. 
Consequently, post-deployment interactions may affect immediate recommendation behavior in an ongoing interaction process \cite{chang2017streaming,song2014online}, but do not naturally translate into persistent improvement of the agent itself. 
As a result, the interaction experiences accumulated after deployment are not systematically incorporated into the learning process. 
Figure \ref{figure1} illustrates this distinction: unlike conventional offline recommendation, where deployment-time interactions remain outside the learning loop, the online paradigm repeatedly updates the recommender system as new interaction data become available.

Bringing post-deployment interactions into the learning loop, however, raises a new question: how should such interactions supervise model updates? A natural alternative is to improve the recommender system using post-deployment interactions through reinforcement learning or other forms of reward-guided optimization \cite{lin2025rec, zhang2025towards, gao2026integrating,kong2026think}. 
However, this approach is intrinsically limited in recommendation settings. 
Unlike tasks for which success can be naturally summarized using a scalar objective, recommendation interactions often contain richer and more structured feedback than a reward function can faithfully preserve \cite{wang2026self,zhang2025survey}. 
The informational value lies not only in whether a recommendation is accepted, but also in how successive user responses reveal and refine preferences.
Once learning is mediated through a reward proxy, optimization is constrained by what the proxy can reliably encode \cite{gao2026integrating}. 
In contrast, abandoning reward supervision does not simplify the problem.
Raw interaction trajectories can be long-horizon,  noisy, and tightly coupled to specific users, candidate sets, or local interaction contexts.
Treating such trajectories as direct optimization targets \cite{li2026chainrec, tian2026astra}, for example through standard supervised fine-tuning, risks fitting session-specific response patterns rather than the transferable knowledge that should govern future recommendation behavior \cite{ye2026online}. 
These limitations motivate a new learning paradigm that can convert post-deployment interactions into reusable supervision, while avoiding both the compression induced by reward proxies and the instability of direct optimization using the raw trajectories.

In this work, we address this challenge through \textbf{Experience}, which captures what should be learned from an interaction trajectory to improve future recommendation behavior. 
Unlike a session summary, an experience abstracts away session-specific details and states a rule that transfers to new contexts.
Yet identifying such experiences is only the first step. 
If the extracted experience remains external to the model, then it can at most support context-dependent adaptation at inference time, rather than persistent improvement of the recommender system. 
To close the learning loop, interaction-derived experiences must be internalized into model parameters, so that the knowledge accumulated from past interactions becomes part of the recommender system’s own capability rather than an auxiliary context that must be repeatedly supplied.
To this end, we propose a novel \textbf{O}nline learning framework--- \textbf{E}xperience-Driven \textbf{A}gentic \textbf{R}ecommendation (\textbf{OEAR}).
OEAR formulates agentic recommendation as a multi-turn interaction process between a RecAgent and a UserAgent. It extracts reusable recommendation-side and user-side experiences from the resulting trajectories, and internalizes them into the agent backbones through on-policy experience distillation, so that it can be retained as persistent agent capability. 
As new interaction data becomes available, the framework repeatedly applies the same cycle, so that improvements in agent behavior can translate into more informative future trajectories, which in turn support further improvement.
Our key contributions can be summarized as:
\begin{itemize}[leftmargin=12pt]
    \item We introduce an online learning framework for agentic recommendation that enables a recommender system to improve using post-deployment interactions, moving beyond static supervision and reward-guided optimization.
    \item We propose an experience-driven learning paradigm that extracts reusable experiences from trajectories, validates each experience by replaying the corresponding decision with and without it, and internalizes the validated experiences into model parameters as persistent recommendation capability.
    \item Extensive experiments on three benchmarks demonstrate that the proposed framework consistently improves agentic recommendation performance.
\end{itemize}

\section{Related Work}

\subsection{Agentic Recommender Systems}
Agentic recommender systems have quickly evolved in recent years. An early line of work focuses on agent-based user simulation \cite{wang2023recagent, zhang2024generative, chen2025recusersim}, where LLM-powered agents are used to simulate user behavior or interaction environments for recommendation. 
A second stage moves from simulation to interactive recommender–user feedback loops \cite{zhang2024agentcf, cai2025agentic}, where the recommender systems and the users are both explicitly modeled and optimized with user interactions. 
A third stage extends agentic recommendation into multi-agent and memory-augmented systems \cite{wang2024macrec, chen2026memrec, liu2025recoworld}, where recommendation is a collaboration between specialized agents or by explicit collaborative memory mechanisms. 
More recent work further explores policy evolution and system-level designs for agentic recommendation \cite{li2026agenticrec, tao2026sager, zhang2026agenticrs}, including end-to-end ranking-oriented recommender agents and self-evolving personalized recommendation policies.
Prior work has progressively enriched how agentic recommenders interact, collaborate, remember, simulate, and reason. 
However, existing studies still focus primarily on the architecture and scaffolding of agentic recommendation, rather than on how post-deployment interactions can be systematically converted into continuous model improvements. 
Our work targets this missing ingredient to support online agentic recommendation using experience extraction and internalization. 

\subsection{Online Learning for Recommendation}
Online learning for recommendation studies how recommender systems can be updated continuously as new interaction data arrives over time. 
An early line of work studies incremental and streaming recommendation, where models are updated continuously from data streams and evaluated using streaming protocols rather than static train-test splits \cite{vinagre2015evaluation, guo2019streaming, xia2023towards}. 
Subsequent work extends this paradigm to session-based and sequential streaming recommendations \cite{zhang2024gpt4rec, guo2019streaming}, emphasizing temporally evolving user behavior and the need to update the recommendation models using continuous interactions. 
More recent studies move closer to continuous recommendations, which explicitly addresses preference drift, stability–plasticity trade-offs, and long-term adaptation in non-stationary recommendation environments \cite{yoo2025continual, lim2025federated}. 
Recent work further broadens the space to practical and deployment-oriented continual recommendation, including federated continuous recommendation, efficient multimodal streaming recommendation, and lightweight adaptation mechanisms for evolving interaction graphs \cite{lim2025federated, qu2025efficient, yoo2024ensuring}. 
In summary, prior online recommendation research has mainly focused on how recommendation models can be incrementally or continuously updated in dynamic data streams. 
In contrast, our work studies online recommendation in an agentic multi-turn setting and improves the model by experience extraction and internalization rather than directly optimizing with streaming interactions alone.  

\begin{figure*}[t]

    \centering

    \includegraphics[width=\textwidth]{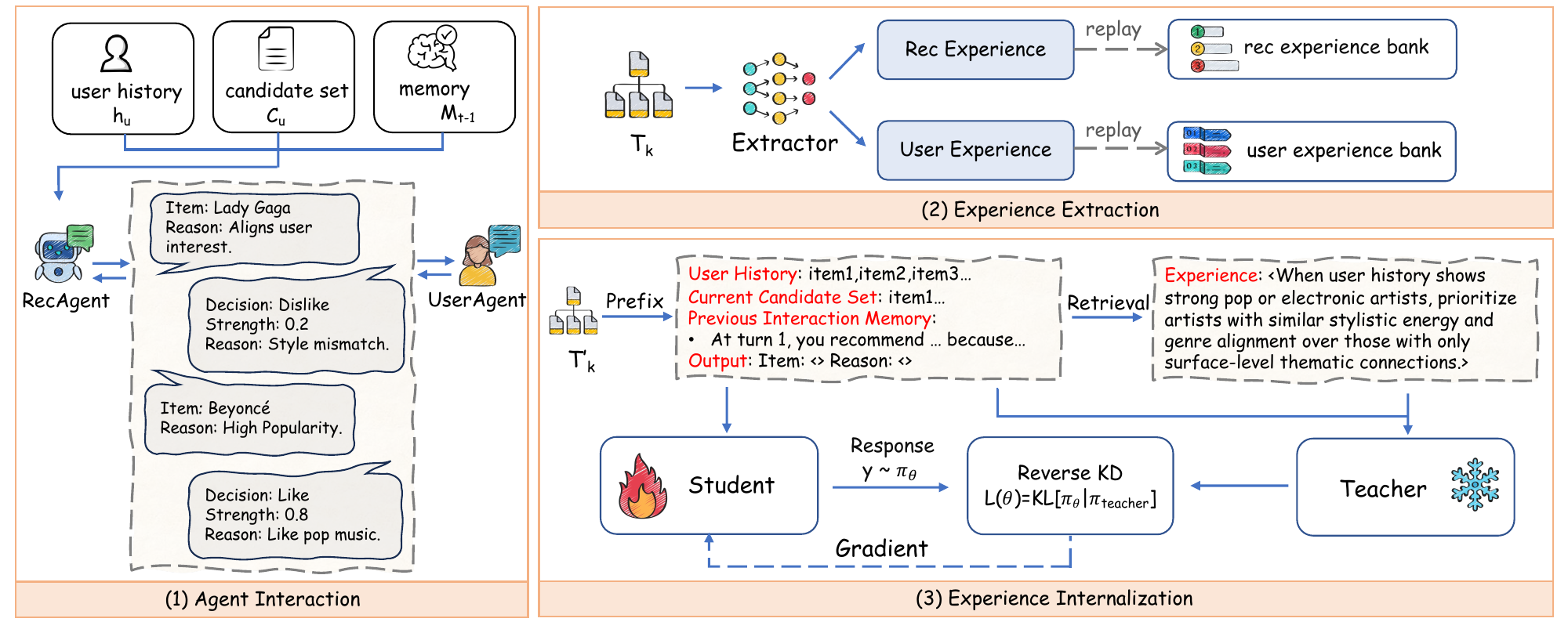}
    \vspace{-9pt}
    \caption{An overview of the proposed OEAR framework.}
    \label{figure2}
    \vspace{-9pt}

\end{figure*}

\section{Preliminaries}
Before presenting our method, we first introduce the basic agentic recommendation setting used throughout this paper. 

\textbf{Architecture}. In our formulation of agentic recommender systems, recommendation is modeled as a multi-turn interactive process between a RecAgent and a UserAgent. 
Unlike conventional one-shot recommendation, RecAgent produces not only a recommendation decision but also an explicit recommendation rationale, while the UserAgent evaluates the recommendation, returns behavioral feedback, and further provides textual justification. 
Such a dual-agent architecture turns recommendation into an iterative loop in which both sides are continuously conditioned on an evolving interaction history.

\textbf{RecAgent.} Formally, let $\mathcal{U}$ denote the user set and $\mathcal{I}$ denote the item set. 
For each user $u \in \mathcal{U}$, let $h_u$ denote the historical interaction sequence for user $u$. 
Suppose that completing one recommendation session for user $u$ requires at most $n$ rounds of interactions between the RecAgent and the UserAgent. 
Let $y_u^*$ denote the target item of the current session.
At turn $t$, the RecAgent takes the user history $h_u$ as input, the candidate set $\mathcal{C}_u \subseteq \mathcal{I}$, and the current memory $\mathcal{M}_{t-1}$, and outputs a recommended item $\hat{y}_t \in \mathcal{C}_u$ together with its recommendation rationale $r_t^{\mathrm{rec}}$. 
More formally, this is:
\begin{equation}
(\hat{y}_t, r_t^{\mathrm{rec}}) = f_{\mathrm{rec}}(h_u, \mathcal{C}_u, \mathcal{M}_{t-1}),
\end{equation}
where $f_{\mathrm{rec}}$ denotes the recommendation engine of the RecAgent with a system prompt (shown in Appendix \ref{appendix}).

\textbf{UserAgent.}
Conditioned on the recommendation pair $(\hat{y}_t, r_t^{\mathrm{rec}})$, the UserAgent outputs a behavioral decision $a_t$, an acceptance strength $s_t$, and a feedback rationale $r_t^{\mathrm{user}}$, or more formally:
\begin{equation}
(a_t, s_t, r_t^{\mathrm{user}}) = f_{\mathrm{user}}(h_u, \hat{y}_t, r_t^{\mathrm{rec}}, \mathcal{M}_{t-1}),
\end{equation}
where $f_{\mathrm{user}}$ denotes the feedback generation engine of the UserAgent. 
Here, $a_t \in \{\texttt{like}, \texttt{dislike}\}$ denotes the user-side behavioral decision, $s_t \in [0,1]$ denotes the corresponding acceptance strength, and $r_t^{\mathrm{user}}$ denotes the textual feedback rationale. 
Importantly, the interpretation of $s_t$ depends on the decision $a_t$: if $a_t=\texttt{like}$, then $s_t \in [0.5,1.0]$, the values closer to $1.0$ indicate stronger acceptance; if $a_t=\texttt{dislike}$, then $s_t \in [0,0.5)$, the values closer to $0$ indicate stronger rejection. 
Using this design, the threshold $0.5$ serves as the polarity boundary between acceptance and rejection, while the magnitude relative to this boundary reflects the intensity of the user-side judgment. 
This richer feedback structure is important for the effectiveness of our framework, as it provides more informative interaction signal for subsequent experience extraction than scalar rewards can alone.

\textbf{Interaction memory.}
The memory $\mathcal{M}_{t-1}$ is a summary of the interaction history accumulated before turn $t$. 
More specifically, it stores the previous recommendation decisions, recommendation rationales, user actions, acceptance strengths, and user feedback rationales from earlier rounds. 
After each turn, the memory is updated with a newly generated interaction tuple so that subsequent recommendation decisions are conditioned not only on the static user history but also on the evolving interaction process, including prior successes, failures, and unresolved preference signals.

\textbf{Multi-turn interaction.}
We denote one interaction turn at round $t$ as:
\begin{equation}
\tau_t = (h_u, \mathcal{C}_u, \mathcal{M}_{t-1}, \hat{y}_t, r_t^{\mathrm{rec}}, a_t, s_t, r_t^{\mathrm{user}}),
\end{equation}
which records the local interaction state and the outputs of the two agents for that round. 
Here, $\tau_t$ captures only the turn-level interaction itself, rather than the full recommendation session.

\textbf{Recommendation trajectory.}
A complete recommendation session for user $u$ is represented as a multi-turn trajectory:
\begin{equation}
\mathcal{T}_u = \big(h_u, y_u^*, \mathcal{C}_u, \{\tau_t\}_{t=1}^{n}, y_u^{\mathrm{final}}),
\end{equation}
where $\{\tau_t\}_{t=1}^{n}$ is the sequence of interaction turns, $y_u^{\mathrm{final}}$ is the final recommended item returned by the system. 
In this way, the trajectory records not only the intermediate interaction process, but also the session-level supervision. The interaction process proceeds iteratively until either the UserAgent outputs a \texttt{like} decision or a maximum number of turns is reached. We set the maximum number as 5.
Once a \texttt{like} decision is produced, the corresponding recommended item is regarded as the final recommendation for the current session. 
If no \texttt{like} decision is observed within the maximum number of interactions allowed, the item recommended in the last turn is used as the final recommendation. Therefore, $y_u^{\mathrm{final}}$ is determined either by the first accepted recommendation or, failing that, by the last recommendation in the trajectory.

\section{Proposed Method}
OEAR is designed to turn post-deployment interactions into persistent improvements of agentic recommender systems. 
It introduces experience as an intermediate form of reusable knowledge, where interaction trajectories are abstracted into transferable experiences and then internalized into agent backbones as persistent capabilities. 
Figure~\ref{figure2} provides an overview of OEAR, and we next introduce its main stages.

\subsection{Experience Extraction}
A complete trajectory records detailed agent interaction behaviors, but many of these details are tied to a particular user, candidate set, or local decision context. 
Therefore, instead of treating the full trajectory as direct supervision, OEAR starts by extracting experience---a compact and reusable piece of interaction-derived knowledge that captures what should be retained from a session to improve future agent behavior. 
Given a complete session trajectory $\mathcal{T}_u$, we extract experience for both agents using an extractor model $f_{\mathrm{ext}}$. 
Instead of performing accumulative extraction by conditioning later extractions on previously extracted experiences, our framework adopts a session-level extraction strategy: each complete trajectory is processed independently, and each session yields one recommendation experience and one user experience.
Formally, let $f_{\mathrm{ext}}$ denote the extractor model. 
For each trajectory $\mathcal{T}_u$, we extract
\begin{equation}
e_u^{\mathrm{rec}} = f_{\mathrm{ext}}(\mathcal{T}_u, p^{\mathrm{rec}}), \qquad
e_u^{\mathrm{user}} = f_{\mathrm{ext}}(\mathcal{T}_u, p^{\mathrm{user}}),
\end{equation}
where $p^{\mathrm{rec}}$ and $p^{\mathrm{user}}$ denote the recommendation-side and user-side extraction prompts, respectively, which are shown in Appendix \ref{appendix}.
Here, $e_u^{\mathrm{rec}}$ is a transferable recommendation experience for the RecAgent, while $e_u^{\mathrm{user}}$ is a transferable judgment experience for the UserAgent.
The two prompts, which are shown in the Appendix, emphasize different extraction objectives: the former focuses on what the recommender should learn from the trajectory, whereas the latter focuses on what the user-side judgment behavior reveals.
In both cases, the extracted experience is constrained to a compact session-level rule rather than a verbose trajectory summary. The extractor model $f_{\mathrm{ext}}$ is the agent model itself: recommendation experiences are extracted by the RecAgent itself, while user experiences are extracted by the UserAgent itself. This yields two raw experience banks, denoted as $\mathcal{E}^{\mathrm{raw}}_{\mathrm{rec}}$ and $\mathcal{E}^{\mathrm{raw}}_{\mathrm{user}}$.


\subsection{Experience Replay}
Experience extraction produces candidate experiences from interaction trajectories, but extraction alone does not guarantee that these experiences are behaviorally useful. 
Because the extracted rules are generated from specific sessions, some may capture transferable decision patterns, while others may reflect noisy feedback, accidental correlations, or overly local context. Therefore, we introduce an \emph{experience replay} stage that validates each raw experience by testing whether it can improve the behavior in the decision scenario from which it was extracted.
Formally, let $e \in \mathcal{E}^{\mathrm{raw}}$ denote a raw extracted experience, and let $c(e)$ denote its original source decision context, i.e., the corresponding original input of the RecAgent and UserAgent. We define a replay function:
\begin{equation}
\mathcal{R}(e, c(e)) \rightarrow s(e),
\end{equation}
which re-injects $e$ into the source context $c(e)$, replays the corresponding agent decision, and returns a replay outcome $s(e)$. Based on this replay outcome, we construct the verified experience bank.

For RecAgent, replay compares when the target item first appears in the iterative recommendation process before and after injecting an experience into the source decision context. Let $y$ denote the recommended item in the first round without the injected experience, let $y^{(e)}$ denote the recommended item in the first round with experience $e$. Let $m$ and $m^{(e)}$ denote the round at which the target item is first recommended before and after experience replay, respectively. We define the recommendation replay score as:
\[
S_{\mathrm{rec}}(e)
=
\left(
1 + 2\,\mathbf{1}\!\left[\min\!\left(m,m^{(e)}\right)=1\right]
\right)
\operatorname{sign}\!\left(m-m^{(e)}\right).
\]
The sign of the score reflects whether the experience causes the target item to appear earlier or later. 
If $m^{(e)} < m$, the target is reached earlier and the score is positive; if $m^{(e)} > m$, it is reached later and the score is negative. 
When the change crosses the first-round recommendation boundary, the score magnitude increases from $1$ to $3$. 
If $m = m^{(e)}$, including $m = m^{(e)} = 1$, the score is $0$, indicating that replay produces no observable improvement under this criterion.
For the UserAgent, let $a$ and $a^{(e)}$ denote the user decisions before and after injecting experience $e$, and let $s$ and $s^{(e)}$ denote their corresponding acceptance strengths. Since \texttt{like} corresponds to $s \geq 0.5$ and \texttt{dislike} to $s < 0.5$, a decision change corresponds to crossing the accept/reject boundary. We define the desirable direction of strength change as:
\begin{equation}
d=
\begin{cases}
+1, & \text{if the recommended item is the target item},\\
-1, & \text{otherwise}.
\end{cases}
\end{equation}
The user replay score is then defined as
\begin{equation}
S_{\mathrm{user}}(e)
=
\left(
1 + 2\,\mathbf{1}\!\left[a \neq a^{(e)}\right]
\right)
d\,
\operatorname{sign}\!\left(s^{(e)} - s\right).
\end{equation}
Here, $d$ determines which direction of strength change is desirable: higher strength is preferred for the target item, whereas lower strength is preferred for a non-target item. When the decision remains unchanged, moving the strength in the desirable direction receives $+1$, moving it in the opposite direction receives $-1$, and no change receives $0$. This applies whether the unchanged decision is correct or incorrect. When the decision changes, the accept/reject boundary is crossed and the score magnitude increases from $1$ to $3$.
Overall, replay serves as a verification stage. The final verified recommendation and user experience banks are therefore defined as:
\begin{equation}
\mathcal{E}^{\mathrm{ver}}_{\mathrm{rec}}
=
\{e_{rec} \in \mathcal{E}^{\mathrm{raw}}_{\mathrm{rec}}
\mid
S_{\mathrm{rec}}(e) > 0\;
\},
\end{equation}
\begin{equation}
\mathcal{E}^{\mathrm{ver}}_{\mathrm{user}}
=
\{e_{user} \in \mathcal{E}^{\mathrm{raw}}_{\mathrm{user}}
\mid
S_{\mathrm{user}}(e) > 0\;
\}.
\end{equation}





\subsection{Experience Internalization}
The goal of this stage is not to keep experience as external context, but to transfer experience-conditioned behavior into model parameters so that the learned knowledge can be internalized and reused without explicitly injecting the same experience at inference time. 
We adopt the on-policy context distillation proposed in \cite{ye2026policy,ye2026online} for experience internalization. 
General knowledge distillation involves two roles: a student model, which is the agent to be improved, and a teacher model, which provides stronger guidance for the student. 
A straightforward approach would be to generate trajectories from the teacher and train the student to imitate them. 
However, such teacher-trajectory distillation may introduce a distribution mismatch between training and inference, because the student is trained on prefixes generated by the teacher but must act under its own generated contexts at inference time. 
To avoid this mismatch, on-policy distillation uses trajectories generated by the student itself. 
Given a student-generated trajectory, we cut it at an intermediate position and treat the preceding tokens as a prefix. 
This prefix is shared as the input of both the student and the teacher, and the teacher's next-token distribution is used as the distillation target to guide the student.

Specifically, for each trajectory $\mathcal{T}_u$, because the RecAgent and the UserAgent act at different stages of the interaction loop in our framework, their prefixes are defined differently.
For the RecAgent, the prefix $x_t^{\mathrm{rec}}$ consists of the user history, the current candidate set, and the interaction memory up to turn $t$, excluding the next recommendation decision. For the UserAgent, the prefix $x_t^{\mathrm{user}}$ additionally includes the current recommendation and its rationale, since both are observed before the user-side judgment is made. The student predicts the next token conditioned only on the on-policy prefix, while the teacher receives the same prefix together with experience and the target item as privileged context. For better understanding, we provide an example in Appendix \ref{appendix} to instantiate the cut of a trajectory.

Importantly, the extracted experience is not internalized on the same trajectory from which it is derived. The source trajectory is used to expose a reusable lesson and to support replay-based validation, but reusing it for distillation may cause the student to imitate session-specific behaviors rather than acquire transferable knowledge. We therefore apply validated experiences to new student-generated trajectories for internalization. This design naturally raises a retrieval problem: given a new internalization trajectory, the framework must decide which validated experience should be provided to the teacher. 
Although replay ensures that each retained experience can improve its own source context, it does not guarantee that the same experience is suitable for new prefixes. 
For example, an experience that corrects popularity-driven bias may be unhelpful for a prefix whose main failure comes from style mismatch or weak preference overlap. 
Therefore, before distillation, we retrieve experiences according to the dominant failure pattern of the current prefix.

Concretely, during replay validation, each retained experience is assigned a coarse correction label annotated by the extractor, indicating the type of error it is most useful for fixing, such as superficial relevance, weak overlap with historical preferences, popularity-driven bias, style mismatch, or noisy interaction signals. 
Similarly, since most internalization trajectories do not identify the target item correctly at the current prefix, we assign each such prefix a failure label describing its dominant error pattern. 
Experience retrieval is then performed by matching the failure label of the current prefix with the correction label of replay-validated experiences. 
In this way, the teacher is enhanced with an experience that is not only behaviorally validated, but also relevant to the current decision context.

%

The distillation objective is defined separately for the RecAgent and the UserAgent. 
For compactness, we first define a token-level reverse Kullback-Leibler (KL) distillation loss. 
Given an on-policy prefix $x$, a retrieved experience $\tilde{e}$, the target item $y^*$, and a student-generated output $o$, we define:
\begin{equation}
\begin{aligned}
\mathcal{D}_{\pi^{\mathrm{stu}},\pi^{\mathrm{tea}}}(x,\tilde{e},y^*,o)
=
\frac{1}{|o|}
\sum_{j=1}^{|o|}
D_{\mathrm{KL}}
\Big(
&\pi^{\mathrm{stu}}(\cdot \mid x,o_{<j}) \\
&\big\|
\pi^{\mathrm{tea}}(\cdot \mid x,\tilde{e},y^*,o_{<j})
\Big),
\end{aligned}
\end{equation}
where $o_{<j}$ denotes the preceding tokens of the student-generated output before position $j$. 
This loss evaluates the teacher and the student on the same student-generated token prefix, while allowing the teacher to additionally access the retrieved experience and the target item as privileged information.

For the RecAgent, let $x_t^{\mathrm{rec}}$ denote an on-policy prefix constructed from a student-generated trajectory in the internalization subset, and let $\tilde{e}^{\mathrm{rec}}$ denote the replay-validated experience retrieved for this prefix. 
The student first generates a recommendation output 
$o^{\mathrm{rec}} \sim \pi_{\mathrm{rec}}^{\mathrm{stu}}(\cdot \mid x_t^{\mathrm{rec}})$ 
conditioned only on the prefix. 
The recommendation-side internalization objective is:
\begin{equation}
\mathcal{L}_{\mathrm{rec}}
=
\mathbb{E}_{x_t^{\mathrm{rec}},\,\tilde{e}^{\mathrm{rec}},\,o^{\mathrm{rec}}}
\left[
S(\tilde{e}^{\mathrm{rec}})
\cdot
\mathcal{D}_{\pi_{\mathrm{rec}}^{\mathrm{stu}},\pi_{\mathrm{rec}}^{\mathrm{tea}}}
(x_t^{\mathrm{rec}},\tilde{e}^{\mathrm{rec}},y^*,o^{\mathrm{rec}})
\right],
\end{equation}
where the expectation is taken over 
$o^{\mathrm{rec}} \sim \pi_{\mathrm{rec}}^{\mathrm{stu}}(\cdot \mid x_t^{\mathrm{rec}})$. 
Here, $\pi_{\mathrm{rec}}^{\mathrm{stu}}$ and $\pi_{\mathrm{rec}}^{\mathrm{tea}}$ denote the student and teacher conditional output distributions of the RecAgent, respectively. Since replay-validated experiences are not equally useful, we directly use the replay score $S(\tilde{e})$ as the weighting coefficient in internalization. As a result, experiences with stronger replay improvement exert stronger supervision during training.

Analogously, for the UserAgent, let $x_t^{\mathrm{user}}$ denote the on-policy prefix before the user-side response, and let $\tilde{e}^{\mathrm{user}}$ denote the replay-validated user-side experience retrieved for this prefix. 
The student generates a user response 
$o^{\mathrm{user}} \sim \pi_{\mathrm{user}}^{\mathrm{stu}}(\cdot \mid x_t^{\mathrm{user}})$ 
conditioned only on the prefix. 
The user-side internalization objective is:
\begin{equation}
\mathcal{L}_{\mathrm{user}}
=
\mathbb{E}_{x_t^{\mathrm{user}},\,\tilde{e}^{\mathrm{user}},\,o^{\mathrm{user}}}
\left[
S(\tilde{e}^{\mathrm{user}})
\cdot
\mathcal{D}_{\pi_{\mathrm{user}}^{\mathrm{stu}},\pi_{\mathrm{user}}^{\mathrm{tea}}}
(x_t^{\mathrm{user}},\tilde{e}^{\mathrm{user}},y^*,o^{\mathrm{user}})
\right],
\end{equation}
where the expectation is taken over 
$o^{\mathrm{user}} \sim \pi_{\mathrm{user}}^{\mathrm{stu}}(\cdot \mid x_t^{\mathrm{user}})$. 
Here, $\pi_{\mathrm{user}}^{\mathrm{stu}}$ and $\pi_{\mathrm{user}}^{\mathrm{tea}}$ denote the student and teacher conditional output distributions of the UserAgent, respectively.

This distillation stage plays the role of experience internalization in OEAR. 
During training, the teacher uses replay-validated and prefix-matched experiences as privileged guidance, while the student only observes the on-policy prefix and its own generated token history. 
By matching the experience-conditioned teacher distribution on student-generated outputs, OEAR transfers experience-guided behavior into the parameters of both the RecAgent and the UserAgent. 
After internalization, the agents can reproduce such behavior without requiring explicit experience input at inference time. 
Over successive online learning rounds, this process progressively converts interaction-derived experience into persistent model capability.

\subsection{Online Learning for Agentic Recommendation}
We formulate OEAR as a sequential interaction-and-update process in which the supervision at each round is generated by the current agents rather than pre-collected in advance. 
At online round $k$, let $D_k$ denote the interaction contexts newly exposed to the agents, and let $\theta_{k-1}$ denote the model parameters obtained after the previous round. 
The current RecAgent and UserAgent interact on $D_k$ to generate trajectories whose distribution depends on the current model state:
\begin{equation}
\mathcal{T}_k \sim p(\mathcal{T}\mid D_k,\theta_{k-1}).
\end{equation}
These trajectories are then used to extract and internalize interaction-derived experiences, producing the updated model:
\begin{equation}
\theta_k
=
\operatorname{Update}
\left(
\theta_{k-1},
\mathcal{T}_k
\right).
\end{equation}
The updated agents are subsequently carried forward to round $k+1$, where they generate new trajectories under $\theta_k$. 
Therefore, the trajectories used for learning in later rounds depend on the updates accumulated from earlier interactions. 
In particular, $\mathcal{T}_{k+1}$ is not a fixed training sample that can be generated independently of round $k$, since changing $\theta_k$ changes the resulting interaction behavior and trajectory distribution.

Within each round, we divide $D_k$ into two disjoint subsets, $D_k^{T}$ and $D_k^{T'}$. 
The current agents interact on $D_k^{T}$ to generate trajectories $\mathcal{T}_k$, from which recommendation-side and user-side experiences are extracted and validated through experience replay. 
They also interact on $D_k^{T'}$ to generate trajectories $\mathcal{T}'_k$, which are converted into prefix-level samples for experience internalization. 
For each prefix, a replay-validated experience is retrieved and provided to the teacher as privileged guidance during on-policy experience distillation.

After distillation, the updated RecAgent and UserAgent are carried forward to the next round. 
Repeating this procedure over $D_1,\ldots,D_K$ forms a closed interaction--learning loop:
\begin{equation}
\theta_{k-1}
\rightarrow
\mathcal{T}_k
\rightarrow
\theta_k
\rightarrow
\mathcal{T}_{k+1}.
\end{equation}
Earlier learning therefore affects not only the current model parameters, but also the trajectories generated for subsequent learning. 
This distinguishes OEAR from offline optimization on a fixed trajectory dataset. 
Although the benchmark interaction contexts are partitioned in advance for reproducibility, the trajectories that provide supervision are generated sequentially by the continually updated agents and are not pre-generated before the corresponding online round.

\section{Experiments}
\subsection{Experimental Settings}
\subsubsection{Datasets.}
We conduct experiments on three recommendation benchmarks, including LastFM~\cite{cantador2011second}, MovieLens~\cite{harper2015movielens}, and Instruments~\cite{hou2024bridging}, which cover different recommendation scenarios. We construct the training and test set following the same setting as prior work~\cite{cai2025agentic}. The statistics of the datasets are summarized in Table~\ref{tab:dataset_stats}. In our online setting, we chronologically partition the training interactions into six consecutive rounds according to interaction timestamps and expose them to the agents sequentially. For each round, the ratio between $D_k^{T}$ and $D_k^{T'}$ is 1:1.

\begin{table}[h]
\centering
\vspace{-9pt}
\caption{Statistics of the datasets.}
\vspace{-9pt}
\label{tab:dataset_stats}
\begin{tabular}{lccc}
\toprule
Dataset & \#Users & \#Items & \#Interactions \\
\midrule
LastFM      & 2,100   & 18,745  & 92,835   \\
ML100k   & 943     & 1,682   & 100,001  \\
Instruments & 903,330 & 112,136 & 1,512,531 \\
\bottomrule
\vspace{-9pt}
\end{tabular}
\end{table}

\subsubsection{Baselines.}
We compare our method with three classes of baselines: (1) Traditional recommendation methods, i.e. SASRec \cite{kang2018self}. (2) Instruction-based agentic recommendation methods, i.e., AFL \cite{cai2025agentic} and iAgent \cite{xu2025iagent}. (3) Reinforcement learning-based agentic recommendation methods, i.e. RecoWorld \cite{liu2025recoworld}.
\subsubsection{Metrics}
Given a user’s interaction history, we first combine the ground-truth item and top 19 items that SASRec predicts, forming a candidate list of 20 items. SASRec is well-trained under the training set. Following prior agentic recommendation settings \cite{cai2025agentic}, the RecAgent recommends one item at each interaction turn and receives user feedback before making the next recommendation. This sequential single-item protocol is widely adopted in interactive agentic recommendation because each recommendation constitutes an action whose feedback conditions the agent's subsequent decision. Then the model’s recommendation ability to identify the correct item is evaluated using HitRatio@1. We evaluate user-feedback simulation with F1 score. We first construct a candidate set consisting of the ground-truth item and top-3 items from SASRec and then evaluate the UserAgent's behavioral decision.
\subsubsection{Hyperparameter Settings}
In this paper, we adopt Qwen3-4B \footnote{\url{https://huggingface.co/Qwen/Qwen3-4B-Instruct-2507}} as the backbone model and train it with LoRA \cite{hu2022lora}. At each round, the agents and their corresponding extractor, replay and labeling models are the same. Moreover, the models used in each round are initialized from the checkpoint trained in the previous round. For both RecAgent and UserAgent training, the batch size is set to 16. Importantly, OEAR performs lightweight updates rather than full retraining. At each online round, the RecAgent is updated for only 100 optimization steps and the UserAgent for 500 steps using LoRA. The learning rate is set to 1e-6 for both agents. 

\subsection{Overall Performance Comparison}
We compare OEAR with all the baselines to examine whether the proposed framework can consistently improve recommendation performance. As shown in Table~\ref{table2}, OEAR achieves the best results across all three datasets. This suggests that the advantage of OEAR does not come merely from introducing interaction into recommendation, but from how interaction-derived information is used for learning. In particular, although RecoWorld also optimizes the RecAgent through interaction, its learning signal is ultimately mediated by reward at the trajectory or outcome level. Such reward-based optimization necessarily summarizes rich interaction traces into a more compressed supervisory signal, making it harder to preserve fine-grained evidence about why a recommendation was relatively good or bad. By contrast, OEAR explicitly extracts, replay-validates, retrieves, and internalizes interaction-derived experience, so that informative decision patterns are retained in a more structured form and can be internalized into the model parameters. \par

For user simulation performance, we only compare with agentic recommendation methods. Table~\ref{table2} shows that OEAR again achieves the best results across all three datasets. This indicates that OEAR improves not only recommendation quality, but also the behavioral fidelity of the user side. In particular, while RecoWorld models the user side primarily as a simulator, OEAR explicitly supports experience internalization for both agents. As a result, OEAR learns a stronger UserAgent and provides more reliable feedback signals during online evolution, which further benefits later rounds of improvement.

\begin{table}[t]
\centering
\caption{Overall performance comparison.}
\vspace{-9pt}
\label{table2}
\small
\setlength{\tabcolsep}{3.5pt}
\begin{tabular}{lcccccc}
\toprule
\multirow{2}{*}{Method} & \multicolumn{3}{c}{Recommendation (HitRatio@1)} & \multicolumn{3}{c}{User Simulation (F1)} \\
\cmidrule(lr){2-4} \cmidrule(lr){5-7}
 & LastFM & ML100k & Instru & LastFM & ML100k & Instru \\
\midrule
SASRec    & 0.0629 & 0.0340 & 0.0400 & --     & --     & --     \\
AFL       & 0.0451 & 0.0404 & 0.0515 & 0.1146 & 0.1440 & 0.1984 \\
iAgent    & 0.0412 & 0.0587 & 0.0890 & 0.1356 & 0.1542 & 0.2642 \\
RecoWorld & 0.1248 & 0.1075 & 0.1854 & 0.1256 & 0.1432 & 0.2254 \\
\midrule
OEAR      & \textbf{0.1900} & \textbf{0.1578} & \textbf{0.2713} & \textbf{0.2084} & \textbf{0.1816} & \textbf{0.2877} \\
\bottomrule
\end{tabular}
\end{table}

\subsection{Effectiveness of Progressive Online Learning}
\begin{figure}[t]

    \centering

    \includegraphics[width=\columnwidth]{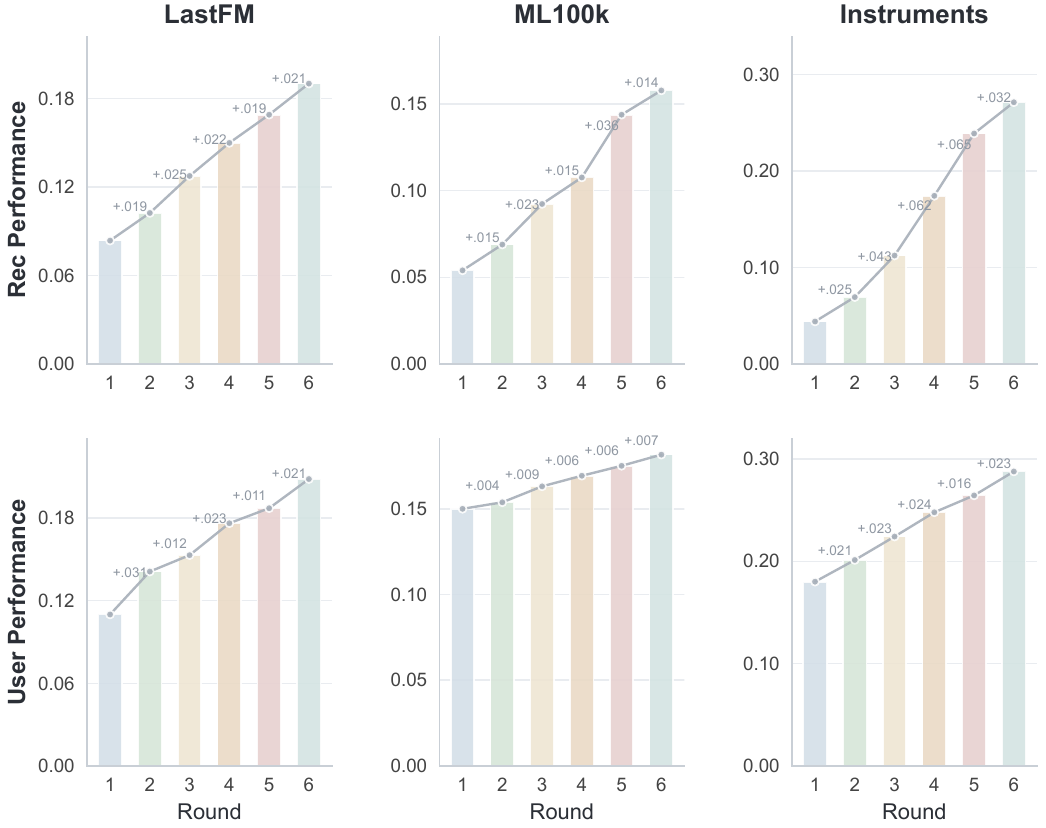}
    \vspace{-9pt}
    \caption{Effectiveness of Progressive Online Learning.}
    \label{figure3}
    \vspace{-9pt}
\end{figure}

To demonstrate the effectiveness of progressive online learning, we present the performance from Round 1 to Round 6 in Figure \ref{figure3}. The figure shows that OEAR improves progressively over successive online learning rounds for both recommendation and user simulation across all three datasets. For the RecAgent, the performance increases steadily from Round 1 to Round 6, indicating that the proposed pipeline can continuously accumulate and internalize useful recommendation-side experience rather than relying only on one-shot adaptation. A similar trend is observed for the UserAgent, whose performance also improves across rounds on all datasets. This is particularly important because it suggests that the framework not only strengthens the recommendation policy, but also continuously refines the user-side decision policy through repeated interaction, replay, retrieval, and distillation. As new data become available in later rounds, the improved agents can in turn generate better trajectories and higher-quality experiences, which provide stronger supervision for subsequent rounds of training.
Overall, the gains across rounds provide direct evidence that OEAR functions as a genuinely progressive online learning framework.

\subsection{Ablation Study}
To examine the role of each major component in OEAR, we design ablation variants for both agents. We first consider \textbf{Target Only}, which removes the retrieved experience from the teacher context while retaining the target item and the same on-policy distillation objective as the full model. This variant isolates the additional contribution of experience beyond target-guided supervision. We further replace retrieval-based matching with random experience sampling (\textbf{Random Experience}) to evaluate the importance of assigning relevant experiences to each prefix. For recommendation-side analysis, we additionally remove UserAgent training (\textbf{w/o User-Training}) and use the initial RecAgent checkpoint (\textbf{Base Rec}). Symmetrically, for user-side analysis, we remove RecAgent training (\textbf{w/o Rec-Training}) and use the initial UserAgent checkpoint (\textbf{Base User}). All variants are compared against the full OEAR model.

\begin{figure}[t]

    \centering

    \includegraphics[width=\columnwidth]{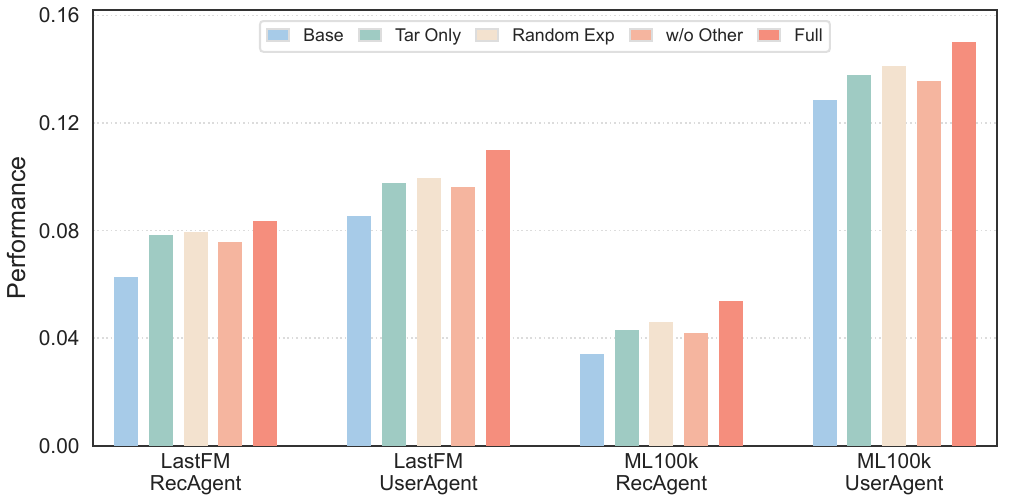}
    \vspace{-9pt}
    \caption{Ablation Study.}
    \label{figure4}
    \vspace{-9pt}
\end{figure}

Due to space constraints, we report the Round-1 ablation results on two representative datasets in Figure~\ref{figure4}. Several observations can be made. First, the full model consistently achieves the best performance, indicating that OEAR benefits from the joint effect of experience guidance, relevance-aware retrieval, and bidirectional agent evolution. Second, the variant with target only consistently underperforms the full model despite using the same target information and on-policy distillation procedure. This result shows that the improvement of OEAR cannot be attributed solely to target-guided supervision; replay-validated experiences provide additional behavioral guidance beyond the target item itself. Third, replacing relevance-aware retrieval with random experience assignment further degrades performance, indicating that experience is most effective when it is matched to the failure pattern of the current prefix. Finally, removing UserAgent training weakens recommendation performance, while removing RecAgent training similarly reduces user simulation quality, demonstrating that the two agents benefit from joint evolution. Overall, these results show that OEAR gains from not only target-guided distillation, but also the complementary contributions of validated experience, relevant experience assignment, and bidirectional agent improvement.

\begin{table}[h]
\centering
\vspace{-9pt}
\caption{Impact of Model Size of Extractor Model.}
\vspace{-9pt}
\label{table3}
\begin{tabular}{lccc}
\toprule
Model & Qwen3-1.7B & Qwen3-4B & Qwen3-8B \\
\midrule
HR      &  0.0815  & \textbf{0.0836} & 0.0767  \\
TR & 0.0916  & \textbf{0.0932} & 0.0895\\

\bottomrule

\end{tabular}
\end{table}
\subsection{Impact of the Model Size of Extractor Model}
To examine the impact of the model size of the extractor model $f_{\mathrm{ext}}$, we fix the policy backbone at Qwen3-4B and vary only the model used for experience extraction and replay. We compare with Qwen-1.7B \footnote{\url{https://huggingface.co/Qwen/Qwen3-1.7B}} and Qwen-8B \footnote{\url{https://huggingface.co/Qwen/Qwen3-8B}}. We show the recommendation performance in Round 1 where we fix the trained UserAgent and the proportion of sessions in which the target item is recommended at any turn (TR) in Table \ref{table3}. 
The results show that the best performance is achieved when the extraction and replay model matches the policy model at the same scale, i.e., Qwen-4B. Under this setting, our original model achieves the best performance on both metrics. Using Qwen-1.7B leads to a moderate degradation on both metrics, while using Qwen-8B causes a larger drop. This suggests that the usefulness of experiential knowledge depends not only on the capability of the extractor model, but also on its compatibility with the policy model that later internalizes such knowledge. In particular, a larger extractor does not necessarily help: although Qwen-8B is stronger in absolute capability, the experiential knowledge it produces may deviate more from the behavior space of the Qwen-4B policy model, making consolidation less effective. Overall, these results support the importance of on-policy consistency: the most effective experiential knowledge is produced by the model that is best aligned with the target policy.

\begin{table}[h]
\centering
\caption{Improvement of Target Recommendation Efficiency.}
\vspace{-9pt}
\label{table4}
\begin{tabular}{llcc}
\toprule
Dataset & Metric & Base & Trained \\
\midrule
\multirow{2}{*}{LastFM}
& Avg. Target Turn & 1.2286 & \textbf{1.1647} \\
& First-Turn TR & 0.0788 & \textbf{0.0804} \\
\midrule
\multirow{2}{*}{ML100k}
& Avg. Target Turn & 1.8200 & \textbf{1.6300} \\
& First-Turn TR & 0.0610 & \textbf{0.0740} \\
\bottomrule
\end{tabular}
\vspace{-9pt}
\end{table}
\subsection{Target Recommendation Efficiency}
To further examine how training changes the recommendation behavior of the RecAgent, we analyze two trajectory-level indicators: the average turn at which the target item is first recommended (Avg. Target Turn), and the proportion of sessions in which the target item is recommended at the first turn (First-Turn TR). We show the performance comparison of RecAgent in Round 1 in Table \ref{table4}. 
Across both datasets, the trained RecAgent consistently reaches the target item earlier and achieves a higher first-turn target recommendation rate. This indicates that the learned policy is not only more likely to eventually reach the correct item, but also more efficient in doing so. In other words, training improves both the speed and the directness of recommendation.
Recommending the target item earlier reduces unnecessary interaction steps, improves the quality of the session from the beginning, and gives the user less chance to drift into uninformative or incorrect interaction paths. Therefore, the increase in first-turn target recommendation, together with the reduction in the target recommendation turn, provides additional evidence that the trained RecAgent has learned a stronger and more decisive recommendation policy.

\subsection{Catastrophic Forgetting Analysis}
To examine whether continual online training causes catastrophic forgetting, we conduct a retention analysis on LastFM using the untrained base RecAgent at Round 0 as the reference model. Specifically, we identify the test sessions in which the Round-0 RecAgent successfully recommends the target item, and define the forgetting rate at Round $r$ as $\mathrm{FR}_r = |\mathcal{C}_0 \setminus \mathcal{C}_r| / |\mathcal{C}_0|$, where $\mathcal{C}_0$ and $\mathcal{C}_r$ denote the sets of successful target-recommendation sessions at Round 0 and Round $r$, respectively. As shown in Table~\ref{table5}, the forgetting rate remains nearly flat across six rounds, fluctuating only slightly rather than exhibiting a cumulative upward trend. Together with the progressive performance improvements observed in Figure~\ref{figure3}, this suggests that OEAR's continual gains are not achieved by progressively overwriting previously successful recommendation behaviors. One possible explanation lies in its on-policy distillation mechanism: the student is trained on prefixes generated under its own policy, while the teacher differs only by access to privileged target and experience information, thereby keeping updates close to the student's own interaction distribution and reducing distribution mismatch. In addition, OEAR retrieves experience relevant to the current prefix, which constrains each update to the behavioral errors exposed by the current interaction rather than broadly shifting the entire policy. These results provide empirical evidence that OEAR supports continual adaptation while maintaining stable previously acquired recommendation behavior.

\begin{table}[h]
\centering
\caption{Forgetting Rate across Online Training Rounds.}
\vspace{-9pt}
\label{table5}
\begin{tabular}{lcccccc}
\toprule
Round & 1 & 2 & 3 & 4 & 5 & 6 \\
\midrule
Forgetting Rate (\%) & 3.29 & 3.76 & 3.29 & 4.23 & 3.76 & 4.23 \\
\bottomrule
\end{tabular}
\vspace{-9pt}
\end{table}
\section{Conclusion}
In this paper, we argued that agentic recommendation requires a new learning paradigm beyond conventional offline supervision and reward-guided optimization. To address this gap, we proposed OEAR, an online framework that treats interaction-derived experience as an intermediate form of reusable supervision. By extracting recommendation and user-side experiences from interaction trajectories and then internalizing them back into the agents, OEAR closes the loop between deployment and learning. Extensive experiments show that OEAR consistently improves both recommendation performance and user-side behavioral modeling, and that these improvements continue over successive online learning rounds. These results suggest that learning from interaction experience is a promising direction for building recommender systems that do not merely act in dynamic environments, but continue to improve within them. More broadly, our framework provides a first step toward recommendation systems in which deployment is not the end of learning, but an ongoing source of reusable knowledge for continual agent evolution.

\section{Ethical Considerations}
This work adheres to established ethical standards for research. All evaluations were conducted on publicly available recommendation datasets. The proposed method is intended to improve recommendation quality through continual agent learning.

\appendix

\section*{Appendix}

\addcontentsline{toc}{section}{Appendix}

\section{Additional Prompts}
\label{appendix}

\begin{tcolorbox}[title={RecAgent Prompt Template},colback=white,colframe=gray!60,colbacktitle=gray!50,coltitle=black,breakable,sharp corners]
You are a recommendation agent.
Given the user history, the current candidate list, and the previous interaction memory in this session, recommend one item from the current candidate list.
Do not recommend an item that has already been recommended.
Use previous user feedback to refine the next recommendation.
The reason should explain the strongest taste or feedback signal behind the choice.

Output format:\\
Item: \textless exact candidate item\textgreater\\
Reason: \textless one short sentence under 20 words\textgreater

\medskip
User history: \{history\}\\
Candidate list: \{candidate\_list\}\\
Previous interaction memory: \{memory\_block\}
\end{tcolorbox}

\begin{tcolorbox}[title={UserAgent Prompt Template},colback=white,colframe=gray!60,colbacktitle=gray!50,coltitle=black,breakable,sharp corners]
You are a music listener. You have previously listened to the following artists.
Judge whether the current recommendation is strong enough to accept.

Important rules:
\begin{itemize}[leftmargin=1.2em,itemsep=0pt,topsep=2pt]
  \item Output ``like'' when the recommendation shows a strong and specific match to the user's core taste and is one of the strongest plausible choices in the current candidate list.
  \item Output ``dislike'' when the recommendation is weak, only surface-level relevant, or clearly dominated by another remaining candidate.
  \item Compare against the current candidate list, but use a substantial-gap standard: output ``dislike'' when another candidate is clearly and meaningfully better, or when the current recommendation lacks a strong and specific taste match.
\end{itemize}

Reason rules:
\begin{itemize}[leftmargin=1.2em,itemsep=0pt,topsep=2pt]
  \item If Decision is like, explain the strong and specific taste match.
  \item If Decision is dislike, explain the weakness or the clearly stronger alternative.
\end{itemize}

Strength rules:
\begin{itemize}[leftmargin=1.2em,itemsep=0pt,topsep=2pt]
  \item 0.85--1.00: a clear strong match that is good enough to confidently accept
  \item 0.60--0.84: a reasonable match with some uncertainty or weaker evidence
  \item 0.30--0.59: only partial or surface-level relevance, not strong enough to accept
  \item 0.00--0.29: a clear mismatch
\end{itemize}

Output format:\\
Decision: \textless like or dislike\textgreater\\
Reason: \textless one short sentence under 20 words\textgreater\\
Strength: \textless float\textgreater

\medskip
User history: \{history\}\\
Candidate list: \{candidate\_list\}\\
Previous interaction memory: \{memory\_block\}\\
Current recommendation: \{rec\_item\}\\
Recommendation reason: \{rec\_reason\}
\end{tcolorbox}

\begin{tcolorbox}[title={Trajectory vs.\ Prefix},colback=white,colframe=gray!60,colbacktitle=gray!50,coltitle=black,breakable,sharp corners]
\textbf{Trajectory:}\\
User history: [A, B, C]\\
Candidate list at turn 1: [X, Y, Z, T]\\
Turn 1: RecAgent recommends X. UserAgent says dislike.\\
Remaining candidates: [Y, Z, T]\\
Turn 2: RecAgent recommends T. UserAgent says like.\\
Session ends.

\medskip
\textbf{Prefix cut from turn 1:}\\
User history: [A, B, C]\\
Current candidate list: [X, Y, Z, T]\\
Previous interaction: None

\medskip
\textbf{Prefix cut from turn 2:}\\
User history: [A, B, C]\\
Current candidate list: [Y, Z, T]\\
Previous interaction:\\
Turn 1: RecAgent recommended X. UserAgent said dislike.
\end{tcolorbox}

\begin{tcolorbox}[title={Rec/User Experience Extractor Prompt Template},colback=white,colframe=gray!60,colbacktitle=gray!50,coltitle=black,breakable,sharp corners]
You are extracting one reusable recommendation/judge experience for a recommendation/user simulation agent.
You will be given one recommendation trajectory from a multi-turn interaction.
Your goal is to summarize one short reusable lesson that can help the recommendation agent make better future recommendations/judgements in similar cases.

Rules:
\begin{itemize}[leftmargin=1.2em,itemsep=0pt,topsep=2pt]
  \item Output exactly one experience item under 35 words.
  \item Do not mention specific item names.
  \item Write only a reusable recommendation lesson for similar future cases.
\end{itemize}
Output format: Experience: \textless one reusable  lesson\textgreater

\medskip
Trajectory: \{TRAJECTORY\}
\end{tcolorbox}

\clearpage
\bibliographystyle{ACM-Reference-Format}
\bibliography{ref}
\end{document}